\documentclass[12pt]{article}

\usepackage{bbm}
\usepackage{epsfig}
\usepackage{array}
\usepackage{float}
\usepackage{dsfont}
\usepackage{amstext}
\usepackage{amsmath}
\usepackage{amssymb}
\usepackage{graphicx}
\usepackage{color}
\usepackage{cancel}
\usepackage{fancybox}
\usepackage{amsmath, amssymb, graphics}
\usepackage{rotating}
\usepackage{a4}
\usepackage{a4wide}

\newcommand{\mathsym}[1]{{}}

\newcommand{\baz}{\begin{array}{cc}}
\newcommand{\bad}{\begin{array}{ccc}}
\newcommand{\bi}{\begin{itemize}}
\newcommand{\ei}{\end{itemize}}
\newcommand{\ba}{\begin{array}{c}}
\newcommand{\ea}{\end{array}}
\newcommand{\D}{\displaystyle}
\newcommand{\dms}{\mbox{$\Delta m^2_{\odot}$}}
\newcommand{\dma}{\mbox{$\Delta m^2_{\rm A}$}}
\newcommand{\meff}{\mbox{$\langle m \rangle$}}

\def\ra{\rightarrow}
\def\be{\begin{equation}}
\def\ee{\end{equation}}
\newcommand{\bea}{\begin{equation} \begin{array}{c}}
\newcommand{\eea}{ \end{array} \end{equation}}

\def\gs{\mathrel{
   \rlap{\raise 0.511ex \hbox{$>$}}{\lower 0.511ex \hbox{$\sim$}}}}
\def\ls{\mathrel{
   \rlap{\raise 0.511ex \hbox{$<$}}{\lower 0.511ex \hbox{$\sim$}}}}

\numberwithin{equation}{section}

\begin{document}

\title{\vspace{-1cm}
\vskip -0.3cm
\vskip 0.4cm
\Large \bf
 Minimal Textures in Seesaw Mass Matrices and their low and
high Energy Phenomenology} 
\author{
Srubabati Goswami$^{a,b}$\thanks{email: \tt sruba@prl.res.in}~,~~
Subrata Khan$^{a}$\thanks{email: \tt subrata@prl.res.in}~,~~
Werner Rodejohann$^c$\thanks{email: \tt
werner.rodejohann@mpi-hd.mpg.de}  \\ \\
{\normalsize \it$^a$Physical Research Laboratory,}\\
{\normalsize \it Navrangpura, Ahmedabad 380009, India}\\ \\
{\normalsize \it$^b$Harish--Chandra Research Institute,}\\
{\normalsize \it Chhatnag Road, Jhunsi, Allahabad 211019, India }\\ \\ 
{\normalsize \it$^c$Max--Planck--Institut f\"ur Kernphysik,}\\
{\normalsize \it  Postfach 103980, D--69029 Heidelberg, Germany} }
\date{}
\maketitle
\thispagestyle{empty}
\vspace{-0.8cm}
\begin{abstract}
\noindent
In an attempt to find minimal scenarios we 
study the implications of Dirac and Majorana 
mass matrices with texture zeros within the type I seesaw mechanism. 
For the Dirac mass matrices we consider 5 zero textures 
which we show to be the most minimal form that can successfully 
account for low energy phenomenology if the Majorana mass matrices are
chosen minimal as well. For those, we consider 
both diagonal and even more minimal non-diagonal forms. 
The latter can be motivated e.g.~by simple $U(1)$ flavour symmetries 
and have two degenerate eigenvalues. We classify the allowed 
textures and discuss the ramifications for 
leptogenesis and lepton flavour violation. 
\end{abstract}

\newpage
\section{Introduction}

Spectacular results from neutrino oscillation experiments in the 
past decade have established the existence of neutrino masses and 
lepton mixing on firm footing. 
However, the smallness of the neutrino masses still continues to be an 
enigma. The most elegant mechanism for generating small neutrino 
masses is the seesaw mechanism in which
one adds heavy right-handed singlets (type I) 
\cite{seesaw1}, scalar triplets (type II) \cite{seesaw2} 
or fermion triplets (type III) \cite{seesaw3} 
to generate small neutrino masses at low scale.
In the context of the type I seesaw mechanism the mass matrix for the 
left handed neutrinos obtained through seesaw diagonalization 
depends on the Dirac type Yukawa coupling matrix of the neutrinos 
($m_D$)  as well as on the bare Majorana mass matrix ($M_R$) of 
the heavy right handed neutrinos:  
\be
m_\nu = -m_D \, M_R^{-1} \, m_D^{\rm T} \, . 
\label{seesaw} 
\ee 
Reconstruction of seesaw mass matrices from low energy observations 
is a challenging task. The main problem is the mismatch in the number of 
parameters 
because 
in general the seesaw framework contains more parameters compared to
what can be obtained from measurements at low energy and
it is not possible to fix the high energy parameters entirely from
low energy data \cite{diffi} (for an overview, see \cite{WR}).   
One possible solution is the appearance of "texture zeros".  
In general, ``zeros'' imply vanishingly small entries in the 
mass matrices whose origin can be traced to flavour symmetries. 
Consideration of texture zeros in the seesaw mass matrices
provides an useful way to handle the problem of parameter 
mismatch since assumption of texture zeros leads to a reduction of 
the number of parameters at high scale 
and thus strengthens the predictive power of the model.  

Implications of two texture zeros in the low energy Majorana mass matrix 
have been studied in \cite{Frampton:2002yf,Tani,mnuzero} and one 
texture zero have been studied in \cite{merle} 
(see also \cite{hybrid} for the case of additional equality of mass
matrix entries). 
Texture zeros in both the charged lepton and neutrino
mass matrices have been studied in \cite{chargedleptonTEX}. 

Within the framework of seesaw mechanism it is often considered more 
natural to consider texture zeros appearing in the Yukawa 
coupling matrix $m_D$ and/or the right-handed
Majorana mass matrix $M_R$ \cite{seesawTEX,Branco:2007nb,CRR}. 
In particular, it has been shown in \cite{Branco:2007nb}
that if $M_R$ is diagonal and if one assumes that all light neutrino 
states are massive 
then the maximal number of zeros that can be accommodated in 
$m_D$ is four. The phenomenology of those cases is studied in detail
in Ref.~\cite{CRR}. 
We will relax this assumption here since the low energy data 
allow one of the neutrino states to be massless. 
In that case, as we show, even with a diagonal $M_R$ one can obtain 
allowed textures consistent with low energy observations 
for a $m_D$ with 5 zeros. 
The results in \cite{Branco:2007nb} were obtained assuming 
both the charged lepton mass matrix and the heavy neutrino 
mass matrix to be diagonal. We also relax the assumption of a 
diagonal $M_R$. 
However, we insist in non-singular $M_R$ in order to avoid light
right-handed (sterile) neutrinos. With this constraint 
we find that the maximal number of zeros that 
is possible for $M_R$ is four and only three such matrices have a  
non-vanishing determinant. 
Interestingly these matrices obey 
$L_\mu - L_\tau$, $L_e - L_\tau$ and $L_e - L_\mu$ flavour symmetry.   
We investigate if  it is possible to have allowed textures 
assuming the above forms for $M_R$ and $m_D$ with 5 zeros, and find
that such textures are disallowed by low energy constraints.   
It is easy to convince oneself that more zeros in $m_D$ are not 
allowed for the minimal forms of $M_R$ that we study \cite{watanabe}. 
To complete the survey we consider the implications for 
Lepton Flavour Violation (LFV) and leptogenesis for the allowed 
textures surviving the constraints from low energy 
phenomenology.\\

The plan of the paper goes as follows: 
in the next Section we discuss the seesaw phenomenology 
including its manifestation for leptogenesis and LFV. 
In Section \ref{sec:ana} 
we perform the texture analysis for $m_D$ with 5 zeros for 
both diagonal $M_R$ (3 zeros) and non-diagonal $M_R$ with 
4 zeros and obeying simple flavour symmetries. The results in terms of 
allowed and disallowed textures are presented in Section 
\ref{sec:res}. 
Apart from low energy phenomenology, 
we also discuss 
the 
implications for leptogenesis 
and LFV  for the allowed solutions. 
Finally we make some comments on Dirac matrices 
with more than 5 zeros and conclude in Section \ref{sec:concl}.

\section{\label{sec:seesaw}Seesaw Phenomenology} 
The low energy neutrino mass matrix 
given in Eq.~(\ref{seesaw}) is symmetric and can in general 
be diagonalized with a unitary matrix as 
\be
V_\nu^{\rm T} \, m_\nu \, V_\nu = D_\nu \, ,
\ee
where $D_\nu$ is diagonal, real and positive and contains the 
mass eigenvalues for the left-handed neutrinos: 
$D_\nu= {\rm diag}(m_1, m_2, m_3)$. 
In the basis in which the charged lepton mass matrix is
real and diagonal, $V_\nu$ coincides with the 
Pontecorvo-Maki-Nakagawa-Sakata (PMNS) 
matrix $U$ and the standard parametrization for this matrix is: 
\begin{equation}
\label{eq:Upara}
 U =  \left(
 \begin{array}{ccc}
 c_{12} \, c_{13} & s_{12}\, c_{13} & s_{13}\, e^{-i \delta}\\
 -c_{23}\, s_{12}-s_{23}\, s_{13}\, c_{12}\, e^{i \delta} &
 c_{23}\, c_{12}-s_{23}\, s_{13}\, s_{12}\,
e^{i \delta} & s_{23}\, c_{13}\\
 s_{23}\, s_{12}-\, c_{23}\, s_{13}\, c_{12}\, e^{i \delta} &
 -s_{23}\, c_{12}-c_{23}\, s_{13}\, s_{12}\,
e^{i \delta} & c_{23}\, c_{13}
 \end{array}
 \right) P \, .
\end{equation}
Here $c_{ij} = \cos \theta_{ij}$, $s_{ij} = \sin \theta_{ij}$,
$\delta$ is the Dirac-type CP-violating phase
and the Majorana phases
$\alpha$ and $\beta$ are contained in the matrix
$P = {\rm diag}(1, e^{i \alpha}, e^{i (\beta + \delta)})$. While all
phases are currently unconstrained, the other mixing parameters are 
determined with increasing precision. Table \ref{tab:data}
summarizes the results from Ref.~\cite{bari}. 

Neutrino oscillation experiments are sensitive to the Dirac 
CP-phase $\delta$. Instead of the phase $\delta$ it is often 
more common to use the rephasing invariant  
quantity $J_{CP}$ (Jarlskog invariant)  
\bea \label{eq:jcp0}
J_{\rm CP} = {\rm Im}
\left\{ U_{e1} \, U_{\mu 2} \, U_{e2}^\ast \, U_{\mu 1}^\ast \right\}
= -\frac{\D {\rm Im} \left\{ h_{12} \, h_{23} \, h_{31} \right\} }
{\D \Delta m^2_{21} \, \Delta m^2_{31} \, \Delta m^2_{32}~}\, , 
\mbox{ where } ~h = m_\nu \, m_\nu^\dagger\,.
\eea
With the parameterization of Eq.~(\ref{eq:Upara}), one has
$J_{\rm CP} = \frac{1}{8} \, \sin 2 \theta_{12}\,
\sin 2 \theta_{23}\, \sin 2 \theta_{13}\,
\cos\theta_{13}\, \sin\delta$.\\ 

\begin{table}[th]
\begin{center}
\begin{tabular}{|c|c|c|c|c|c|} \hline 
 & $\dms/10^{-5}\mathrm{\ eV}^2$ & $\sin^2\theta_{12}$ & $|U_{e3}|$ & $\sin^2\theta_{23}$ &
$\dma/10^{-3}\mathrm{\ eV}^2$ \\ \hline \hline 
Best-fit        &     7.67     &  0.312          &  0.126          &
0.466          &  2.39 \\ \hline 
$1\sigma$ & $7.48 \div 7.83$ & $0.294 \div 0.331$  & $0.077 \div
0.161$  & $0.408 \div 0.539$  & $2.31 \div 2.50$ \\ \hline 
$3\sigma$ & $7.14 \div 8.19$ & $0.263 \div 0.375$  & $<0.214$
& $0.331 \div 0.644$  & $2.06 \div 2.81$ \\ \hline
\end{tabular}
\caption{\label{tab:data}Current best-fit values as well as 1 and
$3\sigma$ ranges of the oscillation parameters \cite{bari}.}
\end{center}
\end{table} 

Another place where CP violation in the lepton sector plays a 
useful role is leptogenesis, even though in general the low and 
high energy CP violations can be completely uncorrelated. 
Both leptogenesis and the later to be discussed Lepton Flavour
Violation (LFV) need to be evaluated in the basis in which $M_R$ is
real and diagonal with $M_1 \le M_2 \le M_3$. 
In general the Majorana mass matrix $M_R$ is non-diagonal
in the basis where the charged current is flavour diagonal.
It is thus written as 
\be
 U_R^\dagger \, M_R \, U_R^* = {\rm diag}(M_1, M_2, M_3) \, .
\label{mrdiagbasis}
\ee
After performing a basis rotation so that the right-handed Majorana
mass matrix $M_R$ becomes diagonal by the unitary matrix $U_R$,  
the Dirac mass matrix $m_D$ gets modified to 
\be
m_D \ra \tilde{m}_D = m_D \,
U_R^{*}  \, .
\ee
Considering the decay of one heavy Majorana neutrino $N_i$ into the
Higgs and lepton doublets, the CP 
asymmetry generated through the interference between tree level and 
one loop heavy Majorana neutrino decay diagrams is given as 
\cite{covi}
\bea
\label{eq:epsIal}
\varepsilon_i^\alpha \D
\equiv \frac{\D \Gamma (N_i \ra \phi \, \bar{l}_\alpha) -
\Gamma (N_i \ra \phi^\dagger \, l_\alpha)}
{\D  \sum\limits_\beta \Big[ \Gamma (N_i \ra \phi \, \bar{l}_\beta) +
       \Gamma (N_i \ra \phi^\dagger \, l_\beta)\Big]}  \\ \D
\,=\, \frac{1}{8 \pi \, v_u^2}
\, \frac{1}{(\tilde m_D^\dagger \, \tilde m_D)_{ii}}  \,
 \sum\limits_{j \neq i}  \, \left(
{\cal I}_{ij}^\alpha \, f(M_j^2/M_i^2) +
{\cal J}_{ij}^\alpha \, \frac{1}{1-M_j^2/M_i^2}
\right)
,
\eea
where
\bea
\label{eq:calIJ}
{\cal I}_{ij}^\alpha = {\rm Im} \Big[ \big(\tilde m_D^\dagger \big)_{i \alpha}
\, \big(\tilde m_D \big)_{\alpha j} \big(\tilde m_D^\dagger \tilde m_D \big)_{ij} \Big]~,~~
{\cal J}_{ij}^\alpha = {\rm Im} \Big[ \big (\tilde m_D^\dagger \big)_{i \alpha}
\, \big(\tilde m_D \big)_{\alpha j} \big(\tilde m_D^\dagger \tilde m_D \big)_{ji} \Big] \,.
\eea
In the MSSM, the function $f(x)$ has the form \cite{covi}
\be
\D f(x) =
\sqrt{x} \, \Big[
\frac{2}{1 - x} - \ln \Big( \frac{1+x}{x} \Big)
 \Big] \,.
\ee
Here $\varepsilon_i^\alpha$ describes
the decay asymmetry of the right-handed neutrino
of mass $M_i$ into leptons of flavour $\alpha = e, \mu, \tau$.
If the rest-mass of the lightest heavy neutrino is much lighter 
than the other two, i.e.~$M_1 \ll M_{2,3}$, 
the lepton asymmetry is dominated by the decay
of this lightest of the heavy neutrinos. In this case $f(M_j^2/M_1^2)
\simeq - 3 \, M_1/M_j$. Moreover, only the first term, proportional to
${\cal I}_{1j}^\alpha$, in Eq.~(\ref{eq:epsIal}) is relevant 
since the second term proportional to ${\cal J}_ {ij}^\alpha$ is
suppressed by an additional power of $M_1/M_j$. Let us mention here
that we will find  all ${\cal J}_{1j}^\alpha$ to be zero for
the successful textures.   
Note furthermore that the
second term in Eq.~(\ref{eq:epsIal}) vanishes when one sums over flavours
to obtain the flavour independent decay asymmetry:
\bea
\label{eq:epsI}
\varepsilon_i \D
\,=\, \sum\limits_\alpha {\varepsilon_i^\alpha}
\equiv
\frac{
\D \sum\limits_\alpha
\Big[
\Gamma (N_i \ra \phi \, \bar{l}_\alpha) -
\Gamma (N_i \ra \phi^\dagger \, l_\alpha)  \Big]}
{\D \sum\limits_\beta \Big[
\Gamma (N_i \ra \phi \, \bar{l}_\beta) +
       \Gamma (N_i \ra \phi^\dagger \, l_\beta)\Big]} \\ 
\,=\, \D
\frac{1}{8 \pi \, v_u^2} \,
\frac{1}{(\tilde m_D^\dagger \, \tilde m_D)_{ii}}
\, \sum\limits_{j \neq i}
{\rm Im} \,
\Big[ \big(\tilde m_D^\dagger \, \tilde m_D \big)^2_{i j}\Big]
\, f(M_j^2/M_i^2)
\\ \D
= \frac{1}{8 \pi \, v_u^2} \,
\frac{1}{(\tilde m_D^\dagger \, \tilde m_D)_{ii}}  \,
 {\cal I}_{ij}
\, , 
\mbox{ where } {\cal I}_{ij} = \sum\limits_\alpha  {\cal
I}_{ij}^\alpha \, .
\eea
The expressions given above for the decay asymmetries are valid for
the MSSM. Their flavour structure is, however,
identical to that of just the Standard Model.\\

Equally important
in leptogenesis are effective mass parameters that are responsible
for the wash-out. With our assumption that a single 
heavy neutrino of mass $M_1$ is relevant for leptogenesis,  
the wash-out of every decay asymmetry $\varepsilon_1^\alpha$ is
governed by an effective mass 
\be
\tilde{m}_1^\alpha =
\frac{(\tilde m_D^\dagger)_{1 \alpha}\, (\tilde m_D)_{\alpha 1}}{M_1}\,.
\ee
The summation of $\tilde{m}_1^\alpha$ over the flavour index $\alpha$
yields $\tilde{m}_1$, which is the relevant parameter for the
wash-out of $\varepsilon_1$. One needs to insert 
the effective masses in the function 
\be
\eta(x) \simeq 
\left( 
\frac{8.25 \times 10^{-3}~{\rm eV}}{x} + 
\left(\frac{x}{2 \times 10^{-4}~{\rm eV}}\right)^{1.16} 
\right)^{-1}\,.
\ee
The final baryon asymmetry is
\cite{flavour_flav0,flavour_flav,flavour_flav1}
\be
Y_B \simeq \left\{
\baz
-0.01 \, \varepsilon_1 \, \eta(\tilde{m}_1)
& \mbox{one-flavour}\, ,\\[0.2cm]
-0.003 
\left(
(\varepsilon_1^e + \varepsilon_1^\mu) \, \eta\left(\frac{417}{589}
(\tilde{m}_1^e + \tilde{m}_1^\mu) \right) +
\varepsilon_1^\tau \, \eta \left(\frac{390}{589}
\tilde{m}_1^\tau \right) \right)
& \mbox{two-flavour}\, ,\\[0.2cm]
-0.003 
\left(
\varepsilon_1^e \, \eta \left(\frac{151}{179}
\, \tilde{m}_1^e  \right) +
\varepsilon_1^\mu \, \eta \left(\frac{344}{537} \, \tilde{m}_1^\mu \right) +
\varepsilon_1^\tau \, \eta \left(\frac{344}{537} \,
\tilde{m}_1^\tau \right) \right) & \mbox{three-flavour}\,.
\ea \right.
\ee
Here 
we have given separate expressions
for one-, two- and three-flavoured
leptogenesis. The three-flavour case occurs for
$M_1 \, (1 + \tan^2 \beta) \le 10^9$ GeV,
the one-flavour case for
$M_1 \, (1 + \tan^2 \beta) \ge 10^{12}$ GeV,
and the two-flavour case (with the tau-flavour decoupling first
and the sum of electron- and muon-flavours, which act indistinguishably)
applies in between.\\

Another possible part of seesaw phenomenology is Lepton Flavour
Violation (LFV) in supersymmetric seesaw scenarios. 
Decays such as $\ell_i \ra \ell_j \gamma$, with flavour indices $i,j$
spanning $(1 = e,\,2 = \mu,\,3 = \tau)$ can depend on the same  
parameters, though in a different form, as the light neutrino masses.  
In mSUGRA scenarios with 
universal boundary conditions for scalar sparticle mass 
matrices, renormalization effects from $M_X$ down to the scale of the 
heavy neutrino masses induce off-diagonal entries in 
the slepton mass matrix \cite{LFV}: 
\be
{\rm BR}(\ell_i \rightarrow \ell_j + \gamma) = c \times 
{\rm BR}(\ell_i \ra \ell_j
\, \nu \overline{\nu}) \,
|(\tilde m_D \, L \, \tilde m_D^\dagger)_{ij}|^2\, ,
\label{eq:branching}
\ee
where the diagonal matrix $L$ is defined as
\be
L_{kl} = \ln \frac{M_X}{M_k} \, \delta_{kl},
\label{eq:L}
\ee
with $M_k$ being the mass of the $k^{\rm th}$ right-handed
neutrino. 
The constant $c$ in the RHS of Eq.~(\ref{eq:branching}) depends
on certain supersymmetry parameters of mSUGRA, specifically
the universal scalar and gaugino masses and the universal
trilinear scalar coupling as well as on $\tan \beta$.
However, we are not interested here in the exact magnitude of the 
branching ratios. We shall instead study the vanishing of certain
branching ratios or the ratio of branching ratios, which is 
independent on the constant.

\section{\label{sec:ana}Texture Analysis}

\subsection{\label{sec:crit}Criteria for successful Candidates and their Phenomenology}

Once we have evaluated the low energy neutrino mass matrix from $m_D$
and $M_R$ containing zero entries, it turns out that invalid cases
can easily be ruled almost at first sight. 

First of all, the rank of $m_\nu$ should be at least two. 
The resulting $m_\nu$ will have zeros in most of the cases. 
A very powerful result obtained in \cite{Frampton:2002yf}
is that low energy Majorana mass matrices with more than
2 zeros are not consistent with 
data\footnote{Let us note here that seesaw 
realizations of low energy two zero textures have been analyzed in
Ref.~\cite{Tani}.}. From the 15 possible two zero 
textures\footnote{If neutrino are Dirac particles then 
up to five zero entries are allowed in $m_\nu$ \cite{HR}.} 
only seven are found to be consistent with the data. 
The allowed patterns are characterized by  
the simultaneous vanishing of the $ee$- and $e\mu$-, the $ee$- 
and $e\tau$-, the $e\mu$- and $\mu\mu$-, 
the $e\tau$- and $\mu\mu$-, the $e\tau$- and $\tau\tau$-, 
the $e\mu$- and $\tau\tau$, and finally the $\mu\mu$- and
$\tau\tau$-entries. The first two possibilities are only possible for
a normal hierarchy, the latter five only for quasi-degenerate
neutrinos. In particular, if e.g.~one neutrino mass is zero 
then it is not allowed that 
$(m_\nu)_{\mu\mu} = (m_\nu)_{\tau\tau} = 0$. 
If the $\mu\tau$ entry of $m_\nu$ is zero, then no other 
entry is allowed to vanish.  
Regarding the presence of only one zero entry in $m_\nu$, 
an important information is that if the determinant of 
$m_\nu$ vanishes, then the $\mu\tau$ element 
can not be zero \cite{merle}. 

Another necessary result to take into account is the ``scaling'' 
behavior of $m_\nu$ \cite{scaling,scal_others,AR}. This denotes the
following form of the mass matrix:
\be \label{eq:scaling}
m_\nu = 
\left( 
\bad 
a & b & b/c \\
\cdot & d & d/c \\ 
\cdot & \cdot & d/c^2
\ea
\right) .
\ee
This form necessarily occurs 
{\it regardless of the form of $M_R$} if the third row of 
$m_D$ multiplied with $c$ is equal to the second row \cite{AR}. 
In the context of texture zeros, this means that 
each of the second and third row of $m_D$ contain only one entry in
the same column. Alternatively, the second row is zero and the third 
row contains only one non-zero element. 
The low energy matrix in Eq.~(\ref{eq:scaling}) is only
compatible with an inverted hierarchy, $m_3 = U_{e3} = 0$ and 
$\tan^2 \theta_{23} = 1/c^2$. Solar neutrino mixing is unspecified by
the above matrix, and the effective mass governing neutrino-less
double beta decay is $\meff = \sqrt{\dma} \, \sqrt{1 - \sin^2 2
\theta_{12} \, \sin^2 \alpha}$. 
There are in principle other variants of
scaling, which will result in $m_\nu$ similar to
Eq.~(\ref{eq:scaling}), but with the first and second, or first and
third column correlated. These are incompatible with data and can be
identified already by the form of $m_D$. 

The simple set of properties summarized here is basically 
all what is needed to distinguish the allowed possibilities 
from the disallowed ones. A somewhat more tricky situation is dealt
with in Section \ref{sec:MRnd}.\\

Let us also comment on the possible impact of the Renormalization Group 
(RG) effects \cite{rgpapers} on the allowed textures. 
The matrix $m_\nu$ that we obtain through seesaw diagonalization is at 
a high scale determined by the masses of the heavy neutrinos. 
Therefore it is expected that the low scale predictions will be 
affected in general by the RG effects. 
However the impact is in general small for hierarchical 
mass spectrum  \footnote{Considerable running even for normal hierarchy 
can be obtained
if running in between the mass
scales of the individual heavy neutrinos is taken into 
account \cite{ssRG}. 
In order to have our conclusions and findings stable against such
effects, we therefore need to assume that the zeros are 
protected by the same symmetry which is responsible for them. 
}.  
It is also well known that the RG effects on the mass matrix
$m_\nu$ are multiplicative in nature. 
Therefore we expect a zero to remain a zero even after RG evolution. 
Thus, an allowed texture is expected to be stable against 
RG corrections \cite{stability}. 
One can also show that mass matrices with the scaling property 
are stable against RG effects \cite{scaling}. 

Finally, we would like to mention that we do not take into account the
fine-tuned possibility that the low energy texture zeros 
are resulting from cancellation of terms.

\subsection{$M_R$ more minimal than diagonal}
Heavy neutrino mass matrices in diagonal form contain three
independent zeros. One can become even more minimal by choosing the
following, non-singular matrices with four independent zeros: 
\begin{eqnarray} \label{eq:MRnd}
M_R^{-1} = 
\begin{pmatrix}
M_1 & 0 & 0\\ 
 0  & 0 & M_2 \\
 0 & M_2 & 0 \\
  \end{pmatrix} ,
\quad  
\begin{pmatrix}
0 & 0 & M_2\\ 
0  & M_1 & 0 \\
M_2 & 0 & 0 \\
  \end{pmatrix} ,\quad  
\begin{pmatrix}
0 & M_2 & 0\\ 
M_2  & 0 & 0 \\
0 & 0 & M_1 \\
  \end{pmatrix}   .
\end{eqnarray}
Interestingly, these correspond to flavour symmetries 
$L_\mu - L_\tau$, $L_e - L_\tau$ and $L_e - L_\mu$, respectively 
\footnote{Note that the
$M_R$ corresponding to the matrices in \ref{eq:MRnd} 
will be obtained by just replacing $M_i$ by $1/M_i$. 
Thus in this convention the 
$M_i$s in  equation \ref{eq:MRnd}
and throughout the text will have inverse mass dimension.}.    
Majorana (and thus symmetric) matrices with 5 or more zeros are all
singular, lead to light sterile neutrinos and will not be studied
here. As we will see, an exactly diagonal $M_R$ leads to 
allowed forms of $m_\nu$ either with 
scaling or with single zero entries. Using the above textures of 
$M_R$ will give typically two simultaneously vanishing entries in 
$m_\nu$.

With $m_D$ being a in general non-symmetric 
$3\times3$ matrix, there are $N =  ^9\!\!C_n$ possibilities to 
place $n$ zero entries in it. We therefore have e.g.~$N = 126, 84$ 
and 36 if $n = 5$, 6 and 7. With our analysis aiming at 5 zero
textures in $m_D$ and with the four forms of $M_R$, there 
are in total 504 candidates. 
Most of them can be ruled out by the arguments 
given in Section \ref{sec:crit}.

\section{\label{sec:res}Results}

In what follows we classify the allowed textures of $m_D$ and $M_R$
in terms of the allowed forms 
of the low energy Majorana mass matrix $m_\nu$. 

\subsection{\label{sec:MRd}Diagonal $M_R$}
If $M_R$ is diagonal then we find in total 18 allowed cases, all of 
which generate a vanishing determinant for $m_\nu$:
\begin{itemize}
\item[(i)] six of them have a vanishing $e\mu$ entry in $m_\nu$. The cases
are listed in Table \ref{tab:12zero}.  
\item[(ii)] another six contain a vanishing $e\tau$ element. 
The cases are listed in Table \ref{tab:13zero}. 
\item[(iii)] the remaining six fulfill the scaling condition in
Eq.~(\ref{eq:scaling}). The cases are listed in Table
\ref{tab:scaling}. 
\end{itemize}

Let us first discuss the neutrino mixing phenomenology. The
predictions of mass matrices obeying scaling were given after 
Eq.~(\ref{eq:scaling}). The other 12 cases are specified by a
vanishing element in $m_\nu$ and $m_1 = 0$ or $m_3 = 0$, depending on
whether a normal or inverted hierarchy is present. 
In case of $(m_\nu)_{e\mu} = 0$ it has been shown that in general 
$\theta_{13}$ must necessarily be non-zero \cite{merle}. 
If $m_1 = 0$ then one finds from the condition 
$(m_\nu)_{e\mu} = 0$ that 
\be \label{eq:12NH}
|U_{e3}| \simeq \frac{m_2 \, \cos \theta_{12}\, \cot \theta_{23}}
{\sqrt{\dma/\sin^2 \theta_{12} - 2 \sqrt{\dma\dms} \, \cos 2 \alpha}}  
\simeq \left\{ \baz 0.077 \div 0.113 & \mbox{ at } 1\sigma  \\ 
0.050 \div 0.144 & \mbox{ at } 3\sigma
\ea \right. \, .
\ee
The numerical range of $|U_{e3}|$ is obtained by  
using the exact expression for $(m_\nu)_{e\mu}$ and varying the 
neutrino parameters in their currently allowed 1 and $3\sigma$ 
ranges. 
In an inverted hierarchy, with $m_3 = 0$ and $m_2 \simeq m_1$ we find 
\be \label{eq:12IH}
|U_{e3}| \simeq \frac{\sin 2 \theta_{12} \, \sin \alpha}{\sqrt{1 - 
\sin^2 2 \theta_{12} \, \sin^2 \alpha}  } \, \cot \theta_{23} \, .
\ee
While in principle $|U_{e3}|$ can be sizable according to this 
equation, it turns out \cite{merle} that the vanishing of the 
$e\mu$-element implies also that $|\sin \alpha| \ll 1$. 
In fact, expanding in terms of $|U_{e3}|$ one finds 
up to first order that 
\bea \D \label{eq:m12}
(m_\nu)_{e\mu} = \left(m_2 \, e^{2 i \alpha} - m_1 \right) 
\cos \theta_{12} \, \sin \theta_{12} \, \cos \theta_{23} \\ \D
+ e^{i \delta} \, \sin \theta_{23} \left( 
e^{2 i \beta} \, m_3 - \sin^2 \theta_{12} \, 
m_2 \, e^{2 i \alpha} - \cos^2 \theta_{12} \, m_1 
\right) \, |U_{e3}|  
\, ,
\eea
We see that for $m_2 \simeq m_1 \gg m_3$ the phase should 
be such that $e^{2 i\alpha} \simeq 1$. This value implies for 
the effective mass that $\meff \simeq 
\sqrt{\dma} \, \cos^2 \theta_{13}$, i.e., there 
are basically no cancellations. We find numerically 
a lower value of $|U_{e3}| \gs 0.005$ when the $3\sigma$ ranges of 
the oscillation parameters are used, whereas for the $1\sigma$ 
ranges we get $|U_{e3}| \gs 0.082$. 
If the $e\tau$-entry of 
$m_\nu$ is zero, then similar expressions as Eqs.~(\ref{eq:12NH}) and
(\ref{eq:12IH}) are obtained, the only change to be made is that 
$\cot \theta_{23}$ should be replaced with $\tan \theta_{23}$, which
has little numerical consequences.\\

Turning to leptogenesis, we see from Tables 
\ref{tab:12zero}, \ref{tab:13zero} and
\ref{tab:scaling} that for all 18 cases there is only 
one physical phase present. In addition, there is always only one 
flavoured ${\cal I}_{1j}^\alpha$ present (which is equal to 
the unflavoured ${\cal I}_{1j}$) and the corresponding 
wash-out parameter $\tilde m_1^\alpha$ (which is equal
to the unflavoured $\tilde m_1$) is non-zero. Note that 
in some cases $m_D$ is such that the 
neutrino with mass $M_1$ decouples. Then we have calculated the 
corresponding quantities for the second neutrino, 
i.e.~${\cal I}_{2j}^\alpha$ and $\tilde m_2^\alpha$. As 
indicated, all ${\cal J}_{1j}^\alpha$ are zero. 
For all 18 possibilities there is the 
potential to accommodate successful leptogenesis.  

We have checked that for the cases with a vanishing $e\mu$ or $e\tau$
element in $m_\nu$ the invariant $J_{\rm CP}$ from 
Eq.~(\ref{eq:jcp0}), 
responsible for low energy CP violation in oscillations is 
non-zero and is 
proportional to the same  phase factor as in ${\cal I}_{ij}$. 
For instance, if we consider  
\be
m_D =  
 \left(
\begin{array}{ccc}
 0 & 0 & {a_3} \\
 0 & {b_2} & 0 \\
 0 & {c_2} & {c_3} \, e^{i {\gamma_3 }}
\end{array}
\right) ,
\ee 
which gives $m_\nu$ with $m_{e \mu} =0$ for a diagonal $M_R$, then the 
only non-zero ${\cal I}_{ij}$ that we get is 
\be
{\cal I}_{23}^\tau = c_2^2 \, c_3^2 \, \sin 2 \gamma_3 \, ,
\ee
to be compared with 
\be
J_{\rm CP} = a_3^2  \, b_2^2  \, c_2^2  \, c_3^2  \, 
(a_3^2  \, (b_2^2 + c_2^2) +
   b_2^2  \, c_3^2)  \, M_2^3  \, M_3^3  \, \sin 2 \gamma_3 \, .
\ee
Thus we see that  
${\cal I}_{ij}^\tau$ and $J_{\rm CP}$ contain the 
same phase factor $\sin 2 \gamma_3$. 
Hence, the Dirac CP phase is always identical to the 
high-energy ``leptogenesis phase''. 
This conclusion is true for all the textures in Tables  
\ref{tab:12zero} and \ref{tab:13zero}. 

An exception is when scaling in $m_\nu$ is present, where there is no
CP violation in oscillation experiments. Here it may be
helpful to evaluate the invariant \cite{utpal}
\[
I_{e \mu} = {\rm Im} \left\{ (m_\nu)_{e e} 
\, (m_\nu)_{\mu\mu} \, (m_\nu)_{e \mu}^\ast \, 
(m_\nu)_{\mu e}^\ast \right\} \, . 
\]
In general, this expression is rather lengthy, but we can express it 
in simple form for the case of $m_3 = 0$ and $U_{e3}=0$. 
Inserting these constraints in the standard parametrization of 
$U$ from Eq.~(\ref{eq:Upara}) and evaluating the 
invariant yields 
{{
$I_{e \mu} = -m_1 \, m_2 \, \dms \, c_{12}^2 \, s_{12}^2 \, c_{23}^4 \, 
\sin 2 \alpha$   in terms of the low energy parameters. 
Similarly $I_{e \tau}$ is also non-zero and depends on the same phase. 
$I_{\mu \tau}$ is however zero. }}
Note that $\alpha$ corresponds to the Majorana phase, the only
physical low energy phase if scaling is present. 
In terms of the parameters of $m_D$ and $M_R$,  
$I_{e \mu}$ can be expressed in terms of the single phase 
that characterizes the Dirac mass matrices. 
For instance if we consider  
\be
m_D = 
\left(
\begin{array}{ccc}
 0 & {a_2} & {a_3} \, e^{i {\alpha_3}} \\
 0 & 0 & {b_3} \\
 0 & 0 & {c_3}
\end{array}
\right) 
\ee
from Table \ref{tab:scaling},  
which gives $m_\nu$ obeying the scaling property, then 
\be
I_{e \mu} = a_2^2 \, a_3^2 \, b_3^4 \, M_2 \, M_3^3 \, \sin 2 \alpha_3
\, . 
\ee
We note from Table \ref{tab:scaling} that the same phase factor 
$\sin 2 \alpha_3$ from $I_{e \mu}$ also appears in 
${\cal I}_{23}^e$. 
Hence, for this case we can say that 
the Majorana CP phase crucial for neutrino-less double beta decay 
is identical to the high-energy ``leptogenesis phase''.
This conclusion is true for all the textures in 
Table \ref{tab:scaling}. Finally, it is worth remarking that the fifth
entry in Table \ref{tab:scaling} corresponds to the results of 
a model based on the flavour symmetry $D_4 \times Z_2$ constructed in
Ref.~\cite{scaling}.

Let us note that in the case of 4 zero 
textures \cite{Branco:2007nb,CRR} there always was an ambiguity in
what regards the one-to-one identification of low and high energy CP
phases with each other. This is not the case for 5 zero textures.\\

Regarding LFV, in case the $e\mu$ entry of $m_\nu$ vanishes, then also
$(m_D \, L \, m_D^\dagger)_{12}$ is zero. Hence, $\mu \ra e \gamma$
will not be detected if the only source of LFV is the supersymmetric
seesaw with mSUGRA conditions studied here. The remaining, in general
non-zero branching ratios turn out to be proportional to the
respective elements of $m_\nu$. Consider the first matrix in Table 
\ref{tab:12zero}. The low energy mass matrix is 
\be
m_\nu = -\left( \bad 
\frac{a_3^2}{M_3} & 0 & \frac{a_3 \, c_3 \, e^{i\gamma_3}}{M_3} \\ 
\cdot & \frac{b_2^2}{M_2} & \frac{b_2 \, c_2}{M_2} \\
\cdot & \cdot & \frac{c_2^2}{M_2} + \frac{c_3^2 \, e^{i\gamma_3}}{M_3}
\ea
\right),
\ee
and for LFV we find 
\be
|(m_D \, L \, m_D^\dagger)_{13}|^2 = a_3^2 \, c_3^2 \, L_3^2 ~,~~ 
|(m_D \, L \, m_D^\dagger)_{23}|^2 = b_2^2 \, c_2^2 \, L_2^2 \, .
\ee
Hence, BR$(\tau \ra e \gamma) \propto |(m_\nu)_{e\tau}|^2$ and 
BR$(\tau \ra \mu \gamma) \propto |(m_\nu)_{\mu\tau}|^2$. The ratio of 
 BR$(\tau \ra e \gamma)$ and BR$(\tau \ra \mu \gamma)$ can however not
be predicted in general because the a priori unknown 
heavy neutrino masses enter this ratio directly and via 
the factors $L_i$. 
The qualitatively same situation is encountered for a vanishing
$e\tau$ entry of $m_\nu$, in which case BR$(\tau \ra e \gamma)$ is
zero and the other BR are proportional to the respective elements of
$m_\nu$, but with different dependence on the heavy masses. 

In contrast, if $m_\nu$ obeys scaling, then no off-diagonal element of
$(m_D \, L \, m_D^\dagger)_{ij}$ vanishes. Consider the first matrix
in Table \ref{tab:scaling}, for which
\be
m_\nu = -\left( \bad
\frac{a_2^2}{M_2} + \frac{a_3^2 \, e^{2 i \alpha_3}}{M_3} & 
\frac{a_3 \, b_3 \, e^{ i \alpha_3}}{M_3} & 
\frac{a_3 \, c_3 \, e^{ i \alpha_3}}{M_3} \\
\cdot & \frac{b_3^2}{M_3} & \frac{b_3 \, c_3 }{M_3} \\ 
\cdot & \cdot &  \frac{c_3^2}{M_3} 
\ea
\right)
\ee
and 
\[ 
|(m_D \, L \, m_D^\dagger)_{12}|^2 = a_3^2 \, b_3^2 \, L_3^2 ~,~~
|(m_D \, L \, m_D^\dagger)_{13}|^2 = a_3^2 \, c_3^2 \, L_3^2 
~,~~
|(m_D \, L \, m_D^\dagger)_{23}|^2 = b_3^2 \, c_3^2 \, L_3^2 \, .
\] 
In particular, we can make the definite prediction 
\be \label{eq:rat1}
\frac{|(m_D \, L \, m_D^\dagger)_{12}|^2}{|(m_D \, L \,
m_D^\dagger)_{13}|^2} = \frac{b_3^2}{c_3^2} = \cot^2 \theta_{23} \, .
\ee
This relation is in fact true in every seesaw model leading to 
scaling \cite{scaling}. A property of the textures is that the
branching ratios are proportional to their respective mass matrix
entries. We can calculate 
\be
\frac{|(m_\nu)_{e\mu}|^2}{|(m_\nu)_{\mu\tau}|^2} 
= \frac{{\rm BR}(\mu \ra e \gamma) \, 
{\rm BR}(\tau \ra \mu \, \nu\overline\nu)
}{{\rm BR}(\tau \ra \mu \gamma)}
\simeq 
\frac{\sin^2 2 \theta_{12} \, \sin^2 \alpha}{\sin^2 \theta_{23} 
\left(1 -  \sin^2 2 \theta_{12} \, \sin^2 \alpha\right)} \, .
\ee
Here, we have neglected $\dms$ with respect to $\dma$. 
This ratio is naturally of order one, but can also vanish. 
Numerically, for the $1\sigma$ 
{{
($3\sigma$) ranges of the oscillation parameter the ratio is 
less than roughly 18.2 (42.9).
If $\alpha = 0$, then 
it becomes $(\dms/\dma)^2 \, \sin^2 2 \theta_{12}/(16 \sin^2
\theta_{23})$.}}
Note that Eq.~(\ref{eq:rat1}) 
implies that $\tau \ra e \gamma$ will be
too rare to become observable. \\

Finally, it is worth mentioning if a $\mu$--$\tau$ symmetric seesaw is
possible. This would imply 
\be
m_D = \left( 
\bad 
a_1 & a_2 & a_2 \\ 
b_1 & b_2 & b_3 \\ 
b_1 & b_2 & b_3 
\ea
\right)~,~~
M_R = \left(
\bad 
M_1 & 0 & 0 \\
0 & M_2 & 0 \\
0 & 0 & M_2 
\ea
\right) .
\ee
This situation would lead to a $\mu$--$\tau$ symmetric 
low energy mass matrix of the form 
\be
m_\nu = \left( 
\bad 
a & b & b \\ 
\cdot & d & f \\ 
\cdot & \cdot & d 
\ea
\right) 
\ee
with $U_{e3} = 0$ and $\sin^2 \theta_{23} = \frac 12$. It can be seen
from Tables \ref{tab:12zero}, \ref{tab:13zero} and \ref{tab:scaling}
that none of the allowed 5 zero textures is compatible with
$\mu$--$\tau$ symmetry. At low energy we can disregard this
possibility at once for the cases of a zero $(m_\nu)_{e \mu}$ 
or a zero $(m_\nu)_{e \tau}$ entry. Namely, it would imply that also 
$(m_\nu)_{e \tau}$ and $(m_\nu)_{e \mu}$ are zero, which is not
allowed by data. Regarding the matrices obeying scaling, there 
can be ``accidental'' $\mu$--$\tau$ symmetry at low energy, 
namely if $\tan^2 \theta_{23} = 1$. However, this is not a consequence
of $\mu$--$\tau$ symmetry of the seesaw mass matrices $m_D$ and
$M_R$.

\subsection{\label{sec:MRnd}Non-diagonal $M_R$}
If $M_R$ takes on the non-diagonal four zero textures 
in Eq.~(\ref{eq:MRnd}) then we can generate two zero textures in
$m_\nu$. However, only four of the seven allowed two zero textures can be 
obtained. For each of the three possibilities $L_e - L_\mu$, 
$L_e - L_\tau$ and $L_\mu - L_\tau$, at first sight 8 potentially 
successful cases survive: 
\bi
\item for every non-{{diagonal}} form of $M_R$ 
there are two cases with a vanishing $ee$ and $e\mu$ entry in
$m_\nu$. 
They have the additional property 
\be \label{eq:add}
|(m_\nu)_{\mu\mu} \, (m_\nu)_{\tau\tau}| = |(m_\nu)_{\mu\tau}^2| 
~\mbox{ and }~
{\rm arg}\left\{(m_\nu)_{\mu\mu} \, (m_\nu)_{\tau\tau} \, 
\left((m_\nu)_{\mu\tau}^\ast \right)^2 \right\} = 0 \, ;
\ee
\item for every non-{{diagonal}} form of $M_R$ 
there are two cases with a vanishing $ee$ and $e\tau$ entry in
$m_\nu$. They have the same property given in Eq.~(\ref{eq:add}); 
\item for every{{ non-diagonal}} form of $M_R$ 
there are two cases with a vanishing $e\mu$ and $\mu\mu$ entry in
$m_\nu$. They have the additional property 
\be 
|(m_\nu)_{ee} \, (m_\nu)_{\tau\tau}| = |(m_\nu)_{e\tau}^2| 
~\mbox{ and }~
{\rm arg}\left\{(m_\nu)_{ee} \, (m_\nu)_{\tau\tau} \, 
\left((m_\nu)_{e\tau}^\ast \right)^2 \right\} = 0 \, ;
\label{c3}
\ee
\item for every {{non-diagonal}} form of $M_R$ 
there are two cases with a vanishing $e\tau$ and $\tau\tau$ entry in
$m_\nu$. They are subject to the condition 
\be 
|(m_\nu)_{ee} \, (m_\nu)_{\mu\mu}| = |(m_\nu)_{e\mu}^2| 
~\mbox{ and }~
{\rm arg}\left\{(m_\nu)_{ee} \, (m_\nu)_{\mu\mu} \, 
\left((m_\nu)_{e\mu}^\ast \right)^2 \right\} = 0 \, ;
\label{c4}
\ee
\ei
All 24 $m_\nu$ have a non-vanishing determinant, hence no neutrino
mass is zero. They also share the property of a zero invariant 
$J_{\rm CP}$, i.e., in the parametrization of the PMNS matrix applied
here the CP phase $\delta$ is zero or $\pi$. 
However, it is known \cite{Frampton:2002yf,mnuzero} that if 
$(m_\nu)_{e\mu} = (m_\nu)_{\mu\mu} = 0$ or if $(m_\nu)_{e\tau} =
(m_\nu)_{\tau\tau} = 0$, the CP phase should be large because the
small ratio of solar and atmospheric $\Delta m^2$ is proportional to
$\cos \delta$. Hence, these 12 cases can be disregarded at once. 

In the cases with $(m_\nu)_{ee} = (m_\nu)_{e\mu} = 0$ 
and $(m_\nu)_{ee} = (m_\nu)_{e\tau} = 0$, it is extremely 
cumbersome to try to obtain analytical estimates from
these equations.
However, we have checked that no point in the allowed 
parameter space can simultaneously satisfy the 
two zero texture conditions and 
conditions (\ref{c3}) and  (\ref{c4}). The reason for this lies in a
clash of the two zero textures which require hierarchical 
neutrino masses, whereas the additional conditions imply 
larger neutrino masses.

\subsection{Dirac Mass Matrices with more than five Zeros}
Finally, in this Subsection we shortly 
discuss Dirac mass matrices with more than five zero entries. 

Among the 9 elements of $m_D$  which are in general complex
the first non-trivial case with maximal 
number of zeros is 8. 
This implies 2 heavy neutrinos
are completely decoupled from the light neutrinos and it is well 
known that one cannot explain the low energy phenomenology 
successfully with only one heavy
neutrino coupled to the system.

If $m_D$ contains seven zeros, then we find no valid $m_\nu$ for the 
four minimal forms of $M_R$ under study. 
{{In fact we have checked that the above statement is true for the 
most general form of $M_R$. }}
One can group the $^7C_9 =
36$ possible forms of $m_D$, in 3 categories. Their respective
structures are  
\begin{eqnarray}
&&m_D = \begin{pmatrix}
0 & 0 & 0\\
 0 & 0 & X \\
0 & 0 & X  \\
  \end{pmatrix} ~,~~ 
\begin{pmatrix}
0 & 0 & 0\\
 0 & 0 & 0 \\
0 & X & X  \\
  \end{pmatrix} ~,~~
\begin{pmatrix}
0 & 0 & 0\\
 0 & X & 0 \\
0 & 0 & X  \\
  \end{pmatrix}  .
  \end{eqnarray} 
The first two matrices will lead, irrespective of $M_R$, 
in the seesaw mechanism to at least two vanishing 
eigenvalues in $m_\nu$ and
are thus ruled out in general. 
The third one can give two non-vanishing eigenvalues but 
we have checked that the final $m_\nu$ contains more than 2 zeros for 
the  most general form of  $M_R$ and hence is disallowed.  

For the case of 6 zeros in $m_D$
the possible types of  patterns are

\begin{eqnarray}
\begin{pmatrix}
0 & 0 & X\\
 0 & 0 & X \\
0 & 0 & X  \\
  \end{pmatrix}~,~~  
 \begin{pmatrix}
X & 0 & 0\\
 0 & X & 0 \\
0 & 0 & X  \\
  \end{pmatrix}~,~~ 
 \begin{pmatrix}
0 & 0 & 0\\
 0 & 0 & 0 \\
X & X & X  \\
  \end{pmatrix}~,~~ 
\begin{pmatrix}
X & 0 & 0\\
 0 & X & 0 \\
X & 0 & 0  \\
  \end{pmatrix} \, .
  \end{eqnarray}

{{For $m_D$'s of the first and third form,  two eigenvalues of 
$m_\nu$ are zero for a general $M_R$. Hence these forms of $m_D$ 
are ruled out. For $m_D$'s of the second type none of the 
eigenvalues of $m_\nu$ are zero and these cannot be ruled out for a
general $M_R$. But we have checked that for the 4 types of 
$M_R$ that we consider here, the resultant $m_\nu$ contains more than 2 zeros
and hence are ruled out. 
For the $m_D$'s of the fourth type one eigenvalue of $m_\nu$ is zero 
but it contains more than 2 zeros in the flavour basis for a general $M_R$ 
and therefore are not allowed.}} 


Thus we conclude that for diagonal $M_R$ the 5 zero textures in $m_D$ 
correspond to scenarios with a minimal 
number of free parameters. In total, we have a seesaw scenario with 8
zero entries, and if the charged leptons are also taken into account
there are in total 14 zeros.

\section{\label{sec:concl}Conclusions and Summary }

In this paper we have investigated the implications of Dirac mass 
matrices ($m_D$) with 5 zeros for seesaw phenomenology. 
For the Majorana mass matrices ($M_R$) 
we consider both diagonal and non-diagonal 
forms. 
The diagonal form contains three zeros. However we noted that more 
minimal forms of $M_R$ containing 4 zeros are possible. 
Three non-singular forms of $M_R$ exist, which correspond to 
$L_\mu -L_\tau$, $L_e - L_\tau$ and $L_e - L_\mu$ symmetry. 
However, with 5 zero textures in $m_D$ they can be shown to be
incompatible with neutrino data. 
We have classified the allowed textures and discussed their 
implications for leptogenesis and lepton flavour violation.  
For $m_D$ with 5 zeros and diagonal $M_R$ there are 18 allowed 
textures 

\begin{itemize}
\item[(i)] 6 patterns with a vanishing $e \mu$ entry; 
\item[(ii)] 6 patterns with a vanishing $e \tau$ entry; 
\item[(iii)] 6 patterns obeying the scaling property with the 
2nd and 3rd columns of $m_\nu$ proportional to each other. 
\end{itemize}

All these 18 cases have one zero mass eigenvalue. 
The phenomenology of the cases (i) and (ii) 
is quite similar:  $\theta_{13}$ is necessarily 
non-zero and sizable (at 3$\sigma$ between 0.05 and 0.14 for the normal 
hierarchy and larger than 0.005 in the inverted hierarchy)  
and $\delta$ is in general non-trivial. 
Both normal and inverted hierarchy is possible.  
Matrices from category (iii) satisfying the scaling property imply 
$U_{e3} = \delta = 0$ and the hierarchy is inverted. Leptogenesis is 
possible for all 18 patterns and for the first 12 cases the 
same phase is responsible for leptogenesis as well as low energy 
CP violation in neutrino oscillations. 
For the 6 allowed textures adhering to the scaling property the 
leptogenesis phase turns out to be the same as the 
Majorana phase responsible for possible cancellations in 
neutrinoless double beta decay. 

We have also considered the manifestations of seesaw mechanism 
for LFV by considering supersymmetric seesaw with mSUGRA conditions 
as the only source for LFV. For the patterns 
with $(m_\nu)_{e \mu}=0$ the 
BR($\mu \rightarrow e \gamma$) turns out to be 
zero. The BR($\tau \rightarrow e \gamma$) and 
BR($\tau \rightarrow \mu \gamma$) are proportional to 
$(m_\nu)_{e\tau}$ and $(m_\nu)_{\mu\tau}$, respectively, but 
depend on the unknown heavy neutrino masses and hence no definite 
predictions are possible for these. Similar conclusions hold true 
for patterns with $(m_\nu)_{e \tau}=0$.  
For mass matrices obeying scaling properties definite predictions for 
LFV in terms of oscillation parameters can be made. 
 

In conclusion, we have shown  
that with the minimal forms of the Majorana mass matrices
that we have
chosen  (non-diagonal with 4 zeros and diagonal corresponding to 3 zeros)  
5 is the maximal number of zeros in Dirac matrices that can
generate successful low energy phenomenology.

\vspace{0.3cm}
\begin{center}
{\bf Acknowledgments}
\end{center}
W.R.~wishes to thank the Physical Research Laboratory in Ahmedabad  
and the organizers of the workshop 
``Astroparticle Physics -- A Pathfinder to New Physics'' in Stockholm, 
where part of this paper was written, for hospitality. 
This work was supported by the ERC under the Starting Grant 
MANITOP and by the Deutsche Forschungsgemeinschaft 
in the Transregio 27 (W.R.). 
S.G.~acknowledges partial support from the Neutrino Project under the
XI-th plan of Harish-Chandra Research Institute.


\newpage

\begin{table}
\begin{center}
\begin{tabular}{|c|c|c|} \hline
$m_D  $ & leptogenesis & wash-out \\ \hline \hline
$ \left(
\begin{array}{ccc}
 0 & 0 & {a_3} \\
 0 & {b_2} & 0 \\
 0 & {c_2} & {c_3} \, e^{i {\gamma_3 }}
\end{array}
\right)$
& ${\cal I}_{23}^\tau = c_2^2 \, c_3^2  \, \sin 2 \gamma_3$  &
$\tilde m_2^\tau = \frac{c_2^2}{M_2} $   \\ \hline
$ \left(
\begin{array}{ccc}
 0 & {a_2} & 0 \\
 0 & 0 & {b_3} \\
 0 & {c_2} & {c_3} \, e^{i {\gamma_3}}
\end{array}
\right)$ & ${\cal I}_{23}^\tau = c_2^2 \, c_3^2  \, \sin 2 \gamma_3$    &   
$\tilde m_2^\tau = \frac{c_2^2}{M_2} $ 
\\ \hline
$ \left(
\begin{array}{ccc}
 a_1 & 0 & 0 \\
 0 & 0 & b_3 \\
 c_1 & 0 & c_3 \, e^{i {\gamma_3}}
\end{array}
\right)$ &  $\ba {\cal I}_{13}^{\tau } = c_1^2 \, c_3^2 \, \sin 2
\gamma_3 
\ea$ & $\tilde{m}_1^{\tau }=\frac{c_1^2}{M_1}$ \\ \hline
$ \left(
\begin{array}{ccc}
 0 & 0 & {a_3} \\
 {b_1} & 0 & 0 \\
 {c_1} & 0 & {c_3} \, e^{i {\gamma_3}}
\end{array}
\right)$ & $ \ba {\cal I}_{13}^{\tau }=c_1^2 \, c_3^2 \, \sin2 \gamma_3 \ea$  
& $\tilde{m}_1^{\tau }=\frac{c_1^2}{M_1}$  \\ \hline
$ \left(
\begin{array}{ccc}
 0 & {a_2} & 0 \\
 {b_1} & 0 & 0 \\
 {c_1} & {c_2} \, e^{i {\gamma_2}} & 0
\end{array}
\right)$ & ${\cal I}_{12}^{\tau } = c_1^2 \, c_2^2 \, \sin 2 \gamma_2$
& $\tilde{m}_1^{\tau } = \frac{c_1^2}{M_1}$  \\ \hline
$ \left(
\begin{array}{ccc}
 {a_1} & 0 & 0 \\
 0 & {b_2} & 0 \\
 {c_1} & {c_2} \, e^{i {\gamma_2}} & 0
\end{array}
\right)$ &  $\ba {\cal I}_{12}^{\tau }=c_1^2 \, c_2^2 \, \sin 2
\gamma_2 \ea$ & $\tilde{m}_1^{\tau }=\frac{c_1^2}{M_1}$  \\ \hline
\end{tabular}
\caption{\label{tab:12zero} The Dirac mass matrix,  
the non-zero expressions relevant for leptogenesis and 
the corresponding wash-out factors. All these cases give  
$m_\nu$ with one vanishing eigenvalue and 
$(m_\nu)_{e \mu} = 0$. The right-handed neutrino
mass matrix is always diagonal. }
\end{center}
\end{table}

\begin{table}
\begin{center}
\begin{tabular}{|c|c|c|} \hline
$m_D$ & Leptogenesis & washout 
\\\hline \hline
$ \left(
\begin{array}{ccc}
 0 & 0 & a_3 \\
 0 & b_2 & b_3 \, e^{i \beta_3} \\
 0 & c_2 & 0
\end{array}
\right)$ 
& ${\cal I}_{23}^\mu = b_2^2 \, b_3^2 \, \sin 2 \beta_3 $ & 
$\tilde m_2^\mu = \frac{b_2^2}{M_2}$ 
\\ \hline
$ \left(
\begin{array}{ccc}
 0 & a_2 & 0 \\
 0 & b_2 & b_3 \, e^{i \beta_3} \\
 0 & 0 & c_3 
\end{array}
\right)$ 
& ${\cal I}_{23}^\mu = b_2^2 \, b_3^2 \, \sin 2 \beta_3 $   & 
$\tilde m_2^\mu = \frac{b_2^2}{M_2}$ 
\\ \hline
$\left(
\begin{array}{ccc}
 0 & 0 & {a_3} \\
 {b_1} & 0 & {b_3} \, e^{i {\beta_3}} \\
 {c_1} & 0 & 0
\end{array} \right)$ 
& $ {\cal I}_{13}^{\mu }= b_1^2 \, b_3^2 \, \sin 2 \beta_3$ & 
$\tilde{m}_1^{\mu }=\frac{b_1^2}{{M_1}}$  \\ \hline
$ \left(
\begin{array}{ccc}
{a_1} & 0 & 0 \\
 {b_1} & 0 & {b_3} \, e^{i {\beta_3}} \\
 0 & 0 & {c_3}
\end{array} \right)$ 
& ${\cal I}_{13}^{\mu }=b_1^2 \, b_3^2 \, \sin 2 \beta_3 $ 
& $\tilde{m}_1^{\mu }=\frac{b_1^2}{{M1}}$  \\ \hline
$ \left(
\begin{array}{ccc}
 0 & {a_2} & 0 \\
 {b_1} & {b_2} \, e^{i {\beta_2}} & 0 \\
 {c_1} & 0 & 0
\end{array} \right)$ 
& ${\cal I}_{12}^{\mu }=b_1^2 \, b_2^2 \, \sin 2 \beta_2$ 
& $\tilde{m}_1^{\mu }=\frac{b_1^2}{{M_1}}$  \\ \hline
$ \left(
\begin{array}{ccc}
 {a_1} & 0 & 0 \\
 {b_1} & {b_2} \, e^{i {\beta_2}} & 0 \\
 0 & {c_2} & 0
\end{array}
\right)$ & $\ba {\cal I}_{12}^{\mu } =b_1^2 \, b_2^2 \, \sin 2 \beta_2
\ea$ & 
$\tilde{m}_1^{\mu }=\frac{b_1^2}{M_1}$  \\\hline
\end{tabular}
\caption{\label{tab:13zero}The Dirac mass matrix,  
the non-zero expressions relevant for leptogenesis and  
the corresponding wash-out factors. All these cases give  
$m_\nu$ with one vanishing eigenvalue and 
$(m_\nu)_{e \tau} = 0$. The right-handed neutrino
mass matrix is always diagonal. }
\end{center}
\end{table}

\begin{table}
\begin{center}
\begin{tabular}{|c|c|c|c|} \hline
$m_D$ & Leptogenesis & washout & $\tan^2 \theta_{23} $
\\\hline \hline
$\left(
\begin{array}{ccc}
 0 & {a_2} & {a_3} \, e^{i {\alpha_3}} \\
 0 & 0 & {b_3} \\
 0 & 0 & {c_3}
\end{array}
\right)$ & 
${\cal I}_{23}^e = a_2^2 \, a_3^2 \, \sin 2 \alpha_3$  &  
$\tilde m_2^e = \frac{a_2^2}{M_2}$ 
& $\frac{c_3^2}{b_3^2}$ \\\hline
$\left(
\begin{array}{ccc}
 0 & {a_2} & {a_3} \, e^{i {\alpha_3}} \\
 0 & {b_2} & 0 \\
 0 & {c_2} & 0
\end{array}
\right)$ & ${\cal I}_{23}^e = a_2^2 \, a_3^2 \, \sin 2 \alpha_3$  
& $\tilde m_2^e = \frac{a_2^2}{M_2}$   & $\frac{c_2^2}{b_2^2}$ \\\hline
$\left(
\begin{array}{ccc}
 {a_1} & 0 & {a_3} \, e^{i {\alpha_3}} \\
 0 & 0 & {b_3} \\
 0 & 0 & {c_3}
\end{array}
\right)$ & $ \ba {\cal I}_{13}^e =a_1^2 \, a_3^2 \, \sin 2 \alpha_3 \ea$ & 
$\tilde{m}_1^e=\frac{a_1^2}{{M_1}}$ & $\frac{c_3^2}{b_3^2}$ \\\hline
$\left(
\begin{array}{ccc}
{a_1} & 0 & {a_3} \, e^{i {\alpha_3}} \\
 {b_1} & 0 & 0 \\
 {c_1} & 0 & 0
\end{array}
\right)$ & $\ba {\cal I}_{13}^e = a_1^2 \, a_3^2 \, \sin 2 \alpha_3 \ea$ & 
$\tilde{m}_1^e=\frac{a_1^2}{{M_1}}$ & $\frac{c_1^2}{b_1^2}$ \\\hline
$\left(
\begin{array}{ccc}
 {a_1} & {a_2} \, e^{i {\alpha_2}} & 0 \\
 0 & {b_2} & 0 \\
 0 & {c_2} & 0
\end{array}
\right)$ & $\ba {\cal I}_{12}^e = a_1^2 \, a_2^2 \, \sin 2 \alpha_2 \ea $ & 
$\tilde{m}_1^e=\frac{a_1^2}{{M_1}}$ & $\frac{c_2^2}{b_2^2}$ \\\hline
$\left(
\begin{array}{ccc}
 {a_1} & {a_2} \, e^{i {\alpha_2}} & 0 \\
 {b_1} & 0 & 0 \\
 {c_1} & 0 & 0
\end{array}
\right)$ & $\ba {\cal I}_{12}^e= a_1^2 \, a_2^2 \, \sin 2 \alpha_2 \ea$ & 
$\tilde{m}_1^e=\frac{a_1^2}{{M_1}}$ & $\frac{c_1^2}{b_1^2}$\\\hline
\end{tabular}
\caption{\label{tab:scaling} The Dirac mass matrix, 
the non-zero expressions relevant for leptogenesis, 
the corresponding wash-out factors 
and the relevant part of the invariant
for CP violation in neutrino oscillations. 
All these cases give 
$m_\nu$ with scaling property between 2nd and 3rd column, i.e., 
an inverted ordering with $m_3 = U_{e3} = 0$. 
The right-handed neutrino
mass matrix is always diagonal.}
\end{center}
\end{table}

\end{document}